\documentstyle[aps,12pt]{revtex}

\begin{document}
\title{Relativistic fluctuating hydrodynamics}
\author{Esteban Calzetta\thanks{%
E-mail address: calzetta@df.uba.ar}}
\address{IAFE and Physics Department, UBA, Buenos Aires, Argentina}
\maketitle

\begin{abstract}
We derive the formulae of fluctuating hydrodynamics appropiate to a
relativistically consistent divergence type theory, obtaining Landau -
Lifshitz fluctuating hydrodynamics as a limiting case.
\end{abstract}

\section{Introduction}

By applying the fluctuation - dissipation theorem\cite{fdt} to the Navier -
Stokes equation, Landau and Lifshitz developed long ago the standard theory
of fluctuating hydrodynamics \cite{ll57} \cite{llfd} \cite{foxuhl}. This
theory is both appealing and successful, and it is consistent with the
fluctuating hydrodynamics which is derived if fluctuation - dissipation
theory is first applied to the Boltzmann equation\cite{fbe}, and then the
hydrodynamic limit is taken by conventional means\cite{hlbe}. It has also
been applied to black hole fluctuations and vacuum decay \cite{pavon}. The
simplest relativistic generalization of Navier - Stokes theory being the
Eckart theory of dissipative fluids\cite{et} (by simplicity, in this paper
we shall not discuss the related Landau - Lifhsitz theory\cite{llfd}, nor
other `first order' theories\cite{hislin85}), it is only natural to provide
a relativistic version of fluctuating hydrodynamics by applying fluctuation
- dissipation theory to it; this step has been taken, and the results
applied to such fields as cosmology\cite{zimdahl}. Trouble is, Eckart theory
has strong drawbacks as a relativistic theory, among them lack of stable
solutions and acausal propagation of perturbations \cite{hislin85} \cite
{hislin83}. This paper asks what changes must be introduced in Landau -
Lifshitz fluctuating hydrodynamics if the Eckart theory is replaced by a
suitable relativistic theory.

Unfortunately, there is not a single causal version of relativistic real
fluid dynamics as compelling as the Eckart and Landau - Lifshitz theories 
\cite{israel87} \cite{gerlin91}, the main contenders being the extended
thermodynamics theories (ETTs) \cite{jou} \cite{isr76} \cite{isrste80} and
the divergence type theories (DTTs) \cite{liu} \cite{gerlin90}, both of
which draw some support from relativistic kinetic theory \cite{isr63} \cite
{isr72} \cite{isrkan} (see \cite{isrste79} for ETTs, \cite{nagreu} for
DTTs). In particular the ETT framework has been extensively applied to
cosmology \cite{ETcosmo}. The relationship between these two approaches is
not yet well understood. For concreteness, we shall adopt the DTT framework,
which is more adapted to a rigorous statement of results.

It ought to be said that the issues raised in this paper are to some extent
academic, since there are strong reasons to believe that in any case the
Eckart type theories are a good phenomenological description of any
relativistic fluid \cite{ger95} \cite{ortreu}\cite{lindblom96}. However, it
is to be expected that small differences will become important when the
theory is pushed to its limits, such as the relativistic theory of
dissipative superfluids (see \cite{isr81} \cite{isr84} \cite{linhis88} for a
ETT approach to this problem, \cite{cnr97} for a DTT one, and \cite{carter}
for a different perspective), the use of relativistic hydrodynamics as a
phenomenological approximation to semiclassical cosmological models \cite
{hu72}, or the emergence of hydrodynamic descriptions as a preferred set of
decoherent histories \cite{hargel} \cite{hucal97}. It is with these
applications in mind that the following considerations were developed.

The next section provides a review of relativistic fluid dynamics; its aim,
of course, is not to substitute the classic introductions such as refs. \cite
{et,israel87,gerlin90}, but rather to have all the relevant results in one
place and notation. We then give a primer on Landau - Lifshitz fluctuation
theory, and develop the subject of fluctuations in DTTs.

\section{Relativistic fluid dynamics}

\subsection{Basic thermodynamics}

Consider a system described by extensive quantities like entropy, energy,
volume, particle number and momentum $S,U,V,N,\vec P$ and intensive ones
like temperature, pressure, chemical potential and velocity $T,p,\mu ,\vec v%
. $

We build a covariant theory by adopting the following rules :

a) Intensive quantities ($T,p,\mu $) are associated to scalars, which
represent the value of the quantity at a given event, as measured by an
observer at rest with respect to the fluid.

b) Extensive quantities ($S,V,N$) are associated to vector currents $%
S^a,u^a,N^a$ (we assume MTW conventions, $c=1$, signature -+++, and latin
indexes go from 0 to 3 \cite{MTW}), such that given a time like surface
element $d\Sigma _a=n_ad\Sigma $, then $-X^ad\Sigma _a$ is the amount of
quantity $X$ within the volume $d\Sigma $ as measured by an observer with
velocity $n^a$. If further the quantity $X$ is conserved, then $X_{;a}^a=0$.
The quantity $u^a$ associated to volume is the fluid four-velocity, and
obeys the additional constraint $u^2=-1$.

c) Energy and momentum are combined into a single extensive quantity and
associated to a tensor current $T^{ab}$. The energy current, properly
speaking, is $U^a=-T^{ab}u_b$.

The entropy current $S^a$ is given by

\[
TS^a=-T^{ab}u_b+pu^a-\mu N^a, 
\]

which we rewrite as

\[
S^a=\Phi ^a-\beta _bT^{ab}-\alpha N^a 
\]

where we introduced the affinity $\alpha =\mu /T$, the thermodynamic
potential $\Phi ^a=p\beta ^a$, and the inverse temperature vector $\beta
^a=T^{-1}u^a$. The Gibbs - Duhem relation becomes

\[
TdS^a=-d(T^{ab}u_b)+pdu^a-\mu dN^a 
\]

We now introduce the concept of a `perfect fluid', as a system whose energy
- momentum tensor takes the form $T^{ab}=\rho u^au^b+p\Delta ^{ab}$, where $%
\Delta ^{ab}=u^au^b+g^{ab}$ (observe that $\rho $ must be the energy density
seen by an observer moving with the fluid). Since we must have $u_adu^a=0$,
we conclude $pdu^a-T^{ab}du_b=0.$ Thus for a perfect fluid, the Gibbs -
Duhem relation reads

\[
dS^a=-\alpha dN^a-\beta _bdT^{ab} 
\]

and

\[
d\Phi ^a=N^ad\alpha +T^{ab}d\beta _b 
\]

which is how the thermodynamic potential got its name. Equivalently

\begin{equation}
\frac{\partial \Phi ^a}{\partial \alpha }=N^a;\qquad \frac{\partial \Phi ^a}{%
\partial \beta _b}=T^{ab}  \label{b}
\end{equation}

The symmetry of the energy momentum tensor implies that the thermodynamic
potential is itself a gradient

\begin{equation}
\Phi ^a=\frac{\partial \chi _p}{\partial \beta _a}  \label{c}
\end{equation}

where $\chi _p$ is the so called generating function. In a covariant theory, 
$\chi _p$ may only depend on the scalars $\alpha $ and $\beta _a\beta ^a$;
working out the corresponding derivatives, we conclude that necessarily $%
N^a=nu^a$, where $n$ is the particle number density seen by a comoving
observer.

We conclude with a word on equilibrium states. Suppose the fluid departs
from equilibrium by a fluctuation $\delta N^a$, $\delta T^{ab}$, consistent
with the conservation laws but otherwise arbitrary. Then the change in
entropy production is

\[
\delta S_{;a}^a=-\alpha _{,a}\delta N^a-\beta _{b;a}\delta T^{ab} 
\]

But for a true equilibrium state the entropy must be stationary, and so we
must have $\alpha _{,a}=\beta _{\left( b;a\right) }=0$ \cite{israel87} (here
and henceforth, brackets stand for symmetrization). Thus the affinity must
be constant, and the inverse temperature vector must be Killing.

\subsection{Eckart's theory of real fluids}

We wish now to describe a weakly dissipative fluid. Following Eckart, we
shall base our description on the same set of variables than for an ideal
fluid, namely, $T^{ab}$ and $N^a$, only now the energy - momentum tensor is
given as

\begin{equation}
T^{ab}=\rho u^au^b+p\Delta ^{ab}+\tau ^{ab}  \label{e1}
\end{equation}

where it is assumed that the viscous stress $\tau ^{ab}$ obeys

\[
\tau ^{ab}u_au_b=0 
\]

and $p$ is defined to be the same function of $\rho $ and $n$ than in the
ideal case. Assuming that the entropy current and thermodynamic potential
are also the same, we get the entropy production rate

\begin{equation}
S_{;a}^a=-\tau ^{ab}\beta _{a;b}  \label{e2}
\end{equation}

Now expand

\[
\beta _{a;b}=\left( \frac 1{T^2}\right) \left[ -T_{,b}u_a+Tu_{a;b}\right] 
\]

\begin{equation}
\tau ^{ab}=u^aq^b+u^bq^a+\pi \Delta ^{ab}+\pi ^{ab}  \label{e3}
\end{equation}

where

\[
u^aq_a=u_a\pi ^{ab}=\pi _a^a=0 
\]

to get

\begin{equation}
S_{;a}^a=\left( \frac{-1}T\right) \left[ \frac{q^a}T\left(
T_{,a}+Tu^bu_{a;b}\right) +\pi u_{;a}^a+\pi ^{ab}u_{a;b}\right]  \label{e4}
\end{equation}

The principle of positive entropy production is satisfied if

\begin{equation}
q^a=-\kappa \Delta ^{ab}\left( T_{,b}+Tu^cu_{b;c}\right)  \label{e5}
\end{equation}

\begin{equation}
\pi =-\zeta _{BV}\;u_{;a}^a  \label{e6}
\end{equation}

\begin{equation}
\pi ^{ab}=-\eta \Delta ^{ac}\Delta ^{bd}\left( u_{c;d}+u_{d;c}-\frac 23%
u_{;e}^eg_{cd}\right)  \label{e7}
\end{equation}

representing heat flux, bulk and shear viscosity, respectively. With these
constitutive relations, energy - momentum conservation yields an
straightforward, covariant generalization of the Navier - Stokes theory.

$\tau ^{ab}$ can be written directly in terms of the covariant derivatives
of the inverse temperature vector as

\begin{equation}
\tau ^{ab}=-B^{abcd}\beta _{\left( c;d\right) }  \label{e8}
\end{equation}

where

\begin{equation}
B^{abcd}=-4\kappa T^2P_V^{abcd}+4\eta TP_{TT}^{abcd}+6\zeta _{BV}TP_S^{abcd}
\label{e9}
\end{equation}

\begin{equation}
P_S^{abcd}=\frac 13\Delta ^{ab}\Delta ^{cd}  \label{e10}
\end{equation}

\begin{equation}
P_V^{abcd}=\left( \frac{-1}2\right) \left[ u^au^c\Delta ^{bd}+u^au^d\Delta
^{bc}+u^bu^c\Delta ^{ad}+u^bu^d\Delta ^{ac}\right]  \label{e11}
\end{equation}

\begin{equation}
P_{TT}^{abcd}=\left( \frac 12\right) \left[ \Delta ^{ac}\Delta ^{bd}+\Delta
^{ad}\Delta ^{bc}-\frac 23\Delta ^{ab}\Delta ^{cd}\right]  \label{e12}
\end{equation}

Observe that $B^{abcd}=B^{cdab}$, and that the $P$'s are actually projection
operators, that is, $P^2=P$ in all three cases, while the product of
different $P$'s vanishes.

Eckart's theory is such a compelling generalization of non relativistic
dissipative hydrodynamics that it is a pity it doesn't work. The resulting
equations allow for non causal propagation, and ipso facto all their
solutions are unstable \cite{hislin85} \cite{hislin83}.

While it is fairly clear that Eckart's theory (and the closely related
Landau's theory as well) must be rejected, it is not clear at all what
should be their replacement. The so - called Israel - Stewart or `second
order' type theories \cite{jou} \cite{isr76} \cite{isrste80} perform much
better with regards to both causality and stability, while keeping much of
the appeal of the Eckart framework, but still lack a rigorous proof of
consistency.

In the following, we shall adopt instead the Geroch - Lindblom `divergence
type' description of a relativistic real fluid \cite{gerlin90}. The
resulting theory is further removed from direct thermodynamic intuition than
the Eckart proposal, but does allow for a rigorous proof of causality and
stability.

\subsection{Divergence type real fluids}

The failure of the Eckart approach to real fluids may be attributed to two
unwarranted assumptions, namely, that the real fluid could be described
within the same set of variables and with the same entropy current than its
perfect counterpart. As a matter of fact, all that equilibrium
thermodynamics suggests is that, whatever extra variables are brought in to
describe the non equilibrium state, they must vanish in equilibrium, and the
entropy current must match its equilibrium value up to first order in the
deviations from equilibrium.

According to Geroch and Lindblom \cite{gerlin90}, description of a
nonequilibrium state requires, besides the particle current and energy
momentum tensor, a new third order tensor $A_{abc}$, obeying an equation of
motion of divergence type

\begin{equation}
A_{bc;a}^a=I_{bc}  \label{dt1}
\end{equation}

and

\[
A_{abc}=A_{acb};\quad A_b^{ab}=0;\;I_a^a=0 
\]

The entropy current is enlarged to read

\begin{equation}
S^a=\Phi ^a-\beta _bT^{ab}-\alpha N^a-A^{abc}\zeta _{bc}  \label{dt2}
\end{equation}

$\zeta _{ab}$ vanishes identically in equilibrium, it is symmetric, and $%
\zeta _a^a=0$. We further require entropy and the thermodynamic potential to
be algebraic functions of their arguments.

The condition that the conservation laws and Eq. (\ref{dt1}) imply positive
entropy production demands that Eqs. (\ref{b}) hold; in particular, the
thermodynamic potential derives from a generating function as in Eq. (\ref{c}%
). The thermodynamic potential is also allowed to depend on the new tensor $%
\zeta _{ab}$; actually

\begin{equation}
\frac{\partial \Phi ^a}{\partial \zeta _{bc}}=A^{abc}  \label{dt3}
\end{equation}

Thus the entropy production

\begin{equation}
S_{;a}^a=-I^{bc}\zeta _{bc}  \label{dt4}
\end{equation}

We obtain positivity by making the former linearly dependent on the latter.

The Eckart theory is actually a particular case of Geroch and Lindblom's\cite
{gerlin90}. Write the generating functional as

\begin{equation}
\chi _E=\chi _p+\left( \frac 12\right) \zeta _{ab}u^au^b  \label{dt5}
\end{equation}

where $\chi _p$ has the same form than for a perfect fluid.

Recalling 
\begin{equation}
\frac{\partial u^a}{\partial \beta _b}=\beta ^{-1}\Delta ^{ab};\quad \frac{%
\partial T}{\partial \beta _a}=T^2u^a  \label{dt5a}
\end{equation}

we get the thermodynamic potential

\begin{equation}
\Phi _E^a=\Phi _p^a+\beta ^{-1}\Delta ^{ab}\zeta _{bc}u^c  \label{dt5b}
\end{equation}

The particle current is the same than for a perfect fluid, and the energy -
momentum tensor is again of the form $T^{ab}=T_p^{ab}+\tau ^{ab}$, where the
dissipative tensor

\begin{equation}
\tau ^{ab}=C^{abcd}\zeta _{cd}  \label{dt6}
\end{equation}

\[
C^{abcd}=T^2\left\{ -P_V^{abcd}+P_{TT}^{abcd}+\frac 16\Delta ^{ab}\left[
4u^cu^d+g^{cd}\right] \right\} 
\]

(in writing this equation, we have used that $\zeta _a^a=0$). Observe that $%
C $ is not symmetric.

This equations allows us to define the viscous stresses in terms of $\zeta
_{ab}$, namely

\begin{eqnarray}
q^a &=&T^2\Delta ^{ab}\zeta _{bc}u^c;\quad \pi =\left( \frac 43\right)
T^2\zeta _{ab}u^au^b;\quad  \nonumber  \label{pitox} \\
\pi ^{ab} &=&T^2\left[ \Delta ^{ac}\Delta ^{bd}+\Delta ^{ad}\Delta ^{bc}-%
\frac 23\Delta ^{ab}\Delta ^{cd}\right] \zeta _{cd}  \label{dt7}
\end{eqnarray}

The new tensor is given by

\begin{equation}
A_E^{abc}=\left( \frac T2\right) \left[ \Delta ^{ab}u^c+\Delta
^{ac}u^b\right]  \label{nt1}
\end{equation}

and its divergence

\begin{eqnarray}
A_{E;a}^{abc} &=&\left( \frac 12\right) \left( T_{,a}+Tu^du_{a;d}\right)
\left[ \Delta ^{ab}u^c+\Delta ^{ac}u^b\right] +  \label{nt2} \\
&&\left( \frac T2\right) \left[ \frac 23u_{;a}^a\left( 4u^bu^c+g^{bc}\right)
+\Delta ^{ab}\left( u_{d;a}+u_{a;d}-\frac 23u_{;e}^e\Delta _{ad}\right)
\Delta ^{dc}\right]  \nonumber
\end{eqnarray}

Observe that we can write

\begin{equation}
A_{E;e}^{ecd}=C^{Tcdab}\beta _{\left( a;b\right) }  \label{nt3}
\end{equation}

where $C^{Tcdab}=C^{abcd}.$

Let us write Eqs. (\ref{dt1}, \ref{dt6} and \ref{nt3}) in shorthand as

\[
\nabla A_E=I 
\]

\[
\tau =C\zeta 
\]

\[
\nabla A_E=C^T\nabla \beta 
\]

These three equations ought to be equivalent to Eq. (\ref{e8})

\[
\tau =-B\nabla \beta 
\]

To obtain this, we must provide a linear relationship

\begin{equation}
I^{ab}=-D^{abcd}\zeta _{cd}  \label{nt4}
\end{equation}

or in shorthand

\begin{equation}
I=-D\zeta  \label{st2}
\end{equation}

where

\begin{equation}
D=C^TB^{-1}C  \label{st3}
\end{equation}

The inverse must exist, since $B$ is positive definite. This equation may be
inverted, to yield

\begin{equation}
B=CD^{-1}C^T  \label{st4}
\end{equation}

where we assume that all matrices have inverses, even the non symmetric ones.

Written in full

\begin{equation}
D^{abcd}=\frac{4T^3}{3\zeta _{BV}}P_{ST}^{abcd}-\frac{T^2}{2\kappa }%
P_V^{abcd}+\frac{T^3}\eta P_{TT}^{abcd}  \label{st5}
\end{equation}

where

\begin{equation}
P_{ST}^{abcd}=\left( \frac 1{12}\right) \left[ g^{ab}+4u^au^b\right] \left[
g^{cd}+4u^cu^d\right]  \label{st6}
\end{equation}

and the other projectors are defined in eqs. (\ref{e11} and \ref{e12}).
Observe that $D$ is a symmetric operator.

This concludes the setting of Eckart's theory in a DTT framework.

\subsection{Causality and divergence type theories}

While the Eckart theory belongs to the DTT class, it is not a `good' theory,
as we shall presently see. To investigate what conditions must a suitable
DTT theory satisfy, we must discuss further the issues of causality and
stability.

The main advantage of the dissipative type theory framework is that the
condition of causality may be expressed in a remarkably simple form. Let us
introduce the symbol $\zeta ^A$ to denote the triad $(\alpha ,\beta ^a,\zeta
^{ab})$, $A_B^a$ the triad $(N^a,T^{ab},A^{abc})$, and $I_B$ the triad $%
(0,0,I_{ab})$. Then the theory is summed up in the equations (cfr. Eqs. (\ref
{b},\ref{dt3},\ref{dt4},\ref{dt1}))

\[
A_B^a=\frac{\partial \Phi ^a}{\partial \zeta ^B} 
\]

\[
S_{;a}^a=-I_B\zeta ^B 
\]

\[
A_{B;a}^a=I_B 
\]

The equations of motion can also be written as

\[
M_{BC}^a\zeta _{;a}^C=I_B 
\]

where

\begin{equation}
M_{BC}^a=M_{CB}^a=\frac{\partial ^2\Phi ^a}{\partial \zeta ^B\partial \zeta
^C}  \label{cdtt1}
\end{equation}

Then the causality condition is that the quadratic form $M_{BC}^aw_a$ be
negative definite for all future directed timelike vectors $w^a$, or,
equivalently, that for any displacement $\delta \zeta ^A$ from an
equilibrium state, the vector $Q^a=M_{BC}^a\delta \zeta ^B\delta \zeta ^C$
be timelike and future oriented \cite{gerlin90}.

In order to see the meaning of this condition, it is interesting to observe
that under any such displacement, the change in the entropy current is
precisely $\delta S^a=-2Q^a$, and therefore the change in entropy density,
as seen by an observer with velocity $v_a$, is $\delta s=2v_aQ^a$. Thus the
causality condition estates that the entropy will be reduced by any
displacement from equilibrium. This goes beyond mere thermodynamic
stability, which only requires entropy reduction at constant energy and
particle number densities.

For imperfect fluids, it is necessary to consider the full quadratic form,
that is, as a functional of $\delta \alpha $, $\delta \beta ^a$ and $\delta
\zeta ^{ab}$. It is easy to see that the Eckart theory cannot possibly meet
the test, since a whole diagonal block is missing. Geroch and Lindblom \cite
{gerlin90} have suggested a simple way of constructing causal theories close
to Eckart's. The idea is to write down a generating functional of the form
(cfr. Eq. (\ref{dt5}))

\begin{equation}
\chi =\chi _E+\gamma \left( T\right) Q\left( u^a,\zeta ^{ab}\right)
\label{tct1}
\end{equation}

where $Q$ is a positive definite scalar quadratic form on the $\zeta ^{ab}$%
.Then the thermodynamic potential

\begin{equation}
\Phi ^a=\Phi _E^a+u^aQT^2\frac{d\gamma }{dT}+\Delta ^{ab}T\gamma \frac{%
\partial Q}{\partial u^b}  \label{tct2}
\end{equation}

and, provided

\begin{equation}
\frac{d\gamma }{dT}\gg \left| \frac \gamma T\right| >0,  \label{tct3}
\end{equation}

the second term dominates and guarantees the positivity of the corresponding
vector $Q^a$. In what follows we shall assume one such extension of Eckart's
theory has been adopted.

This closes our review of causal relativistic hydrodynamics. We now proceed
to our main subject, namely, the form the fluctuation - dissipation theorem
takes in this context.

\section{Fluctuation theory}

\subsection{The classical fluctuation - dissipation theorem}

The simplest possible application of fluctuation - dissipation theory
relates to an homogeneous system described by time - dependent variables $%
x^i $ \cite{llsp} \cite{foxuhl}. Let

\[
X_i=-\frac{\delta S}{\delta x^i} 
\]

Then the entropy production rate is given by

\[
\dot S=-X_i\dot x^i 
\]

Following common usage, we refer to the $X$'s as thermodynamic forces, and
the $\dot x$'s as thermodynamic fluxes. Then the principle of positive
entropy production on average is satisfied by posing a linear relationship

\begin{equation}
\dot x^i(t)=-\int dt^{\prime }\;\gamma ^{ij}(t,t^{\prime })X_j(t^{\prime
})+\xi ^i(t)  \label{lange1}
\end{equation}

where $\gamma $ is positive definite (Onsager's reciprocity principle
further asserts that $\gamma ^{ij}=\pm \gamma ^{ji}$ according to whether $%
x^i$ behaves as $x^j$ under time reversal or not, and we also assume
causality). Under equilibrium, we have the average

\begin{equation}
\left\langle x^i(t)X_j(t^{\prime })\right\rangle =\frac{\delta x^i(t)}{%
\delta x^j(t^{\prime })}  \label{einstein}
\end{equation}

which follows from Einstein's formula relating entropy to fluctuations, and
averages are time traslation invariant. Therefore

\[
\left\langle x^j\dot x^i\right\rangle +\left\langle x^i\dot x^j\right\rangle
=0 
\]

If the noise $\xi $ is Gaussian, then

\[
\left\langle x^j\xi ^i\right\rangle =\int dt^{\prime }\;\left\langle \xi
^i\xi ^k(t^{\prime })\right\rangle \frac{\delta x^j(t)}{\delta \xi
^k(t^{\prime })} 
\]

From the equations of motion, we conclude that for a given initial condition 
$x^j(t^{\prime })$, the solution is written in terms of a single Green
function as

\[
x^i(t)=G_j^i(t,t^{\prime })x^j(t^{\prime })+\int_{t^{\prime }}^td\tau
\;G_j^i(t,\tau )\xi ^j(\tau ) 
\]

where $G_j^i(t^{+},t)=\delta _j^i$. Therefore

\[
\frac{\delta x^i(t)}{\delta \xi ^j(t^{\prime })}=\frac{\delta x^i(t)}{\delta
x^j(t^{\prime })} 
\]

(which expresses in symbols Onsager's insight that the regression of
microscopic fluctuations follows the same rules than macroscopic ones)

Thus

\begin{equation}
\left\langle \xi ^i\xi ^j\right\rangle =\left( \frac 12\right) \left[ \gamma
^{ij}+\gamma ^{ji}\right]  \label{noise1}
\end{equation}

which is our basic result.

The relationship of forces to fluxes may be inverted to yield

\begin{equation}
X_i=G_{ij}\dot x^j+\theta _i  \label{lange2}
\end{equation}

Then it is immediate that

\begin{equation}
\left\langle \theta _i(t)\theta _j(t^{\prime })\right\rangle =\left( \frac 12%
\right) \left[ G_{ij}+G_{ji}\right]  \label{noise2}
\end{equation}

We may also wish to introduce new fluxes $y=M\dot x$, conjugated to forces $%
Y=M^{-1}X$. The relationship between fluxes and forces becomes

\[
y=RY+\sigma ;\quad R=M\gamma M:\quad \sigma =M\xi 
\]

Then

\[
\left\langle \sigma \sigma \right\rangle =R 
\]

This generalized fluctuation dissipation theorem will be relevant in what
follows.

We shall now apply this basic framework to Eckart's theory and to a causal
relativistic theory.

\subsection{Fluctuations in Eckart's theory}

In order to derive the spontaneous fluctuations of a Eckart type real fluid,
we first regard a fluid theory as an instance of the above example, with the
index $i$ now running over a continuous as well as a discrete range. We also
go back to the entropy production rate Eq. (\ref{e2}) and identify $\tau
^{ab}$ as a `flux' multiplying the corresponding `force' $\beta _{(a;b)}$
(which, as it should, vanishes in equilibrium) (since we shall presently
derive these results from a DTT viewpoint, we only give here the somewhat
sketchy original Landau argument \cite{ll57}; the derivation is spelled out
in full in Fox and Uhlembeck\cite{foxuhl}). We then postulate a Langevin -
type relationship \cite{llfd}

\begin{equation}
\tau ^{ab}(x)=-\int d^4x^{\prime }\sqrt{-g(x^{\prime })}B^{abc^{\prime
}d^{\prime }}\delta ^{(4)}(x,x^{\prime })\beta _{(c^{\prime };d^{\prime
})}(x^{\prime })+s^{ab}(x)  \label{fet1}
\end{equation}

where $\delta ^{(4)}(x,x^{\prime })$ is the covariant four dimensional delta
function. The deterministic part is chosen to reproduce the Eckart
constitutive relationships.

Since the coefficients are already symmetric, the fluctuation dissipation
theorem suggests

\begin{equation}
\left\langle s^{ab}s^{cd}\right\rangle =B^{abcd}\delta ^{(4)}(x,x^{\prime })
\label{fet3}
\end{equation}

On the other hand, if we decompose the stochastic energy - momentum tensor
as in Eq. (\ref{e3})

\begin{equation}
s^{ab}=u^aq_s^b+u^bq_s^a+\pi _s\Delta ^{ab}+\pi _s^{ab}  \label{fet4}
\end{equation}

we find

\[
\left\langle q_s^aq_s^c\right\rangle =2\kappa T^2\Delta ^{ac}\delta
^{(4)}\left( x,x^{\prime }\right) 
\]

\[
\left\langle \pi _s^{ab}\pi _s^{cd}\right\rangle =2T\eta \left[ \Delta
^{ac}\Delta ^{bd}+\Delta ^{ad}\Delta ^{bc}-\frac 23\Delta ^{ab}\Delta
^{cd}\right] \delta ^{(4)}\left( x,x^{\prime }\right) 
\]

\[
\left\langle \pi _s\pi _s\right\rangle =2T\zeta _{BV}\delta ^{(4)}\left(
x,x^{\prime }\right) 
\]

\[
\left\langle \pi _s\pi _s^{ab}\right\rangle =0 
\]

\begin{equation}
\left\langle q_s^a\left( \pi _s\Delta ^{cd}+\pi _s^{cd}\right) \right\rangle
=0  \label{fet5}
\end{equation}

Which generalize the Landau - Lifshitz formulae to the relativistic regime 
\cite{llfd} \cite{zimdahl}.

We are finally prepared to consider the issue of fluctuations in a
relativistically consistent, divergence type theory.

\subsection{ Fluctuations in divergence type theories}

In a divergence type theory, the entropy flux takes the form (cfr. Eq. (\ref
{dt2}))

\[
S^a=\Phi ^a-\zeta ^BA_B^a 
\]

The definition

\[
A_B^a=\frac{\partial \Phi ^a}{\partial \zeta ^B} 
\]

implies the entropy production

\[
S_{;a}^a=-\zeta ^BA_{B;a}^a 
\]

The Langevin - type equations of motion read

\begin{equation}
A_{B;a}^a=I_B+j_B  \label{ecmov}
\end{equation}

Where $j_B$ is the stochastic noise. So far, all formulae in this section
have been ultralocal, meaning that $\Phi ^a$ and $I_B$ are algebraic
functions of the field variables at the same point (as opposed, e. g., to
depending on their derivatives).

A satisfactory theory must predict vanishing mean entropy production in
equilibrium, namely

\begin{equation}
\left\langle \zeta ^B(x)I_B(x)\right\rangle +\left\langle \zeta
^B(x)j_B(x)\right\rangle =0  \label{limitecoinc}
\end{equation}

However, because the coincidence limit of the correlation functions in the
left hand side may not be well defined, we must impose this condition in a
smeared form, or else, coupling it to an elementary causality consideration,
request the stronger condition

\begin{equation}
\left\langle \zeta ^B(x)I_B(x^{\prime })\right\rangle +\left\langle \zeta
^B(x)j_B(x^{\prime })\right\rangle =0  \label{sinlimitecoinc}
\end{equation}

for every spacelike pair $(x,x^{\prime })$.

In the linear approximation, the first term yields

\begin{equation}
\left\langle \zeta ^B(x)I_B(x^{\prime })\right\rangle =I_{(B,C)}\left\langle
\zeta ^B(x)\zeta ^C(x^{\prime })\right\rangle  \label{aaa}
\end{equation}

We now appeal to the general theory above to make the assumption that the
noise is Gaussian, and, since the dynamics is purely local, {\it white, }%
meaning that

\begin{equation}
\left\langle j_B(x)j_C(x^{\prime })\right\rangle =\sigma _{BC}\delta
^{(4)}(x,x^{\prime })  \label{whiteness}
\end{equation}

where $\sigma _{BC}$ is a given matrix. Under these hypothesis we obtain

\[
\left\langle \zeta ^B(x)j_B(x^{\prime })\right\rangle =\sigma _{BC}\frac{%
\partial \zeta ^B(x)}{\partial j_C(x^{\prime })} 
\]

Our goal is to derive the matrix $\sigma _{BC}$, and our result shall be
that indeed $\sigma _{BC}=-I_{(B,C)}.$ To arrive to this end, we must show
that, at equal times, 
\begin{equation}
\left\langle \zeta ^B(x)\zeta ^C(x^{\prime })\right\rangle =\frac{\partial
\zeta ^B(x)}{\partial j_C(x^{\prime })}  \label{susbst}
\end{equation}

Let us multiply both sides of this equation by the non singular matrix

\[
M_{AB}=n_a\frac{\partial ^2\Phi ^a}{\partial \zeta ^A\partial \zeta ^B} 
\]

Where $n_a$ is the unit normal to some Cauchy surface containing both $x$
and $x^{\prime }$ (henceforth, {\it the surface}). In the linear
approximation 
\[
M_{AB}\zeta ^B=\frac{\partial S^0}{\partial \zeta ^A} 
\]

where $S^0=-n_aS^a$ is the entropy density in an adapted coordinate system.
In equilibrium, we may apply Einstein's formula (cfr. Eq. (\ref{einstein}))
above, to conclude

\begin{equation}
M_{AB}\left\langle \zeta ^B(x)\zeta ^C(x^{\prime })\right\rangle =-\delta
_A^C\delta (x,x^{\prime })  \label{lema1}
\end{equation}

where $\delta (x,x^{\prime })$ is the three dimensional covariant delta
function on the Cauchy surface. On the other hand 
\begin{equation}
M_{AB}\frac{\partial \zeta ^B(x)}{\partial j_C(x^{\prime })}=-\frac{\partial
A_A^0(x)}{\partial j_C(x^{\prime })}  \label{lema1b}
\end{equation}

where $A_A^0(x)=-n_aA_A^a(x)$. Let us write the equations of motion Eq. (\ref
{ecmov}) as

\[
\frac{\partial A_A^0(x)}{\partial t}+L=j_A(x) 
\]

where $L$ involves the field variables on the surface, but not their normal
derivatives, and $\partial /\partial t\equiv n^a\partial /\partial x^a.$
Indeed

\begin{equation}
\frac{\partial A_A^0(x)}{\partial j_C(x^{\prime })}=\delta _A^C\delta
(x,x^{\prime })  \label{lema2}
\end{equation}

(at equal times). From Eqs. (\ref{lema1}), (\ref{lema1b}) and (\ref{lema2}),
we obtain the desired result

\begin{equation}
\left\langle j_Aj_B\right\rangle =\left( \frac{-1}2\right) \left[ \frac{%
\delta I_A}{\delta \zeta ^B}+\frac{\delta I_B}{\delta \zeta ^A}\right]
\label{lanDTT4}
\end{equation}

Eqs. (\ref{ecmov}) and (\ref{lanDTT4}) are the basic equations incorporating
fluctuations to the DTT framework, and are our main result.

While our derivation does not apply to Eckart's theory, since the matrix $M$
above fails to be non singular, we may consider it as a limiting case of
causal DTTs and still apply the final result. Since $I_\alpha =I_{\beta
^a}=0 $, and in equilibrium $I_{\zeta ,\alpha }=I_{\zeta ,\beta }=0$, we get
the reassuring result that particle number and energy momentum conservation
are not violated. For the remaining equation we have, in the shorthand
notation of the previous section

\begin{eqnarray*}
C^T\nabla \beta &=&\nabla A=I+j=-D\zeta +j \\
\left\langle jj\right\rangle &=&D
\end{eqnarray*}

The first equation is an algebraic equation for the variables $\zeta ^{ab}$.
Its solution reads $\zeta =\zeta _{old}+\zeta _s$, where $\zeta _{old}$
corresponds to the usual expression in terms of the gradients in temperature
and velocity (which may themselves be stochastic) and $\zeta _s=D^{-1}j$ is
a purely stochastic component with self correlation $D^{-1}$. For the random
viscous stresses we find (cfr. Eq. (\ref{dt6}))

\[
s=C\zeta _s 
\]

and therefore

\[
\left\langle ss\right\rangle =CD^{-1}C^T 
\]

But according to Eq. (\ref{st4}), $CD^{-1}C^T\equiv B$, and these are just
the Landau - Lifshitz fluctuation formulae Eq. (\ref{fet3}).

This deceptively simple derivation should not disguise the fact that the
Eckart theory is pathological. The fluctuation formulae Eqs. (\ref{ecmov})
and (\ref{lanDTT4}) can only be expected to yield sensible results in the
context of truly causal theories, like those described in Eqs. (\ref{tct1}, 
\ref{tct2} and \ref{tct3}).

\subsection{Fluctuations in causal theories}

We finally arrive at our designated goal, namely, the characterization of
hydrodynamic fluctuations in a truly consistent relativistic dissipative
theory. The problem is that we do not have a unique generalization of
Eckart's theory to rely on, but rather a wide family of seemingly plausible
extensions. We shall therefore appeal to Occam's razor, and concentrate on
the simplest possible example, namely, the family of generalizations of
Eckart's theory introduced in equations (\ref{tct1},\ref{tct2} and \ref{tct3}%
) above. While we shall appeal to rather drastic approximations to keep
things simple, it should be clear that this serves illustration purposes
only, and does not affect the validity of the noise correlation formula Eq. (%
\ref{lanDTT4}) above.

A DTT theory is defined by its generating functional and the driving forces $%
I^{ab}$ in the equations of motion Eq. (\ref{dt1}). Since the shortcomings
of Eckart's theory concern only the first, it is simplest to assume that the
driving forces of the full theory remain the same, namely, they are still
given by Eqs. (\ref{nt4} and \ref{st5}) above. The generating functional is
modified by adding a new term depending on a positive definite scalar $Q$
(cfr. eq. (\ref{tct1})) and a function $\gamma $ of temperature. The
simplest choice for the latter is a scale - free power law, $\gamma \sim T^n$%
. Since the restriction eq. (\ref{tct3}) excludes the exponent $n=1$, we
shall settle for the next available alternative, $n=2$.

Concerning $Q$, we notice that the Eckart's theory already contains such an
scalar, namely the entropy production rate $-I^{ab}\zeta _{ab}$, and it is
simplest to assume that both are related. We therefore seek a $Q$ scalar
with the structure (cfr. eq. (\ref{st5}), and note that this $Q$ depends
explicitly on $T$)

\begin{equation}
Q\sim \left( \frac 14\right) \left[ \frac{4\tau _{BV}}{3\zeta _{BV}}\left(
\zeta P_{ST}\zeta \right) -\frac{\tau _{HC}}{\kappa T}\left( \zeta P_V\zeta
\right) +\frac{\tau _{SV}}\eta \left( \zeta P_{TT}\zeta \right) \right]
\label{fs0}
\end{equation}

The physical meaning of the new $\tau $ coefficients will become clear soon;
for the time being we only remark that all three are assumed to be positive.
As a matter of fact, using the Eckart's theory results as a guide, we know
that in the end, contrasting eqs. (\ref{dt6}) and (\ref{dt7}) with eqs. (\ref
{e5}, \ref{e6} and \ref{e7}), we shall obtain $P_{ST}\zeta \sim \zeta _{BV}$%
, $P_V\zeta \sim \kappa T$ and $P_{TT}\zeta \sim \eta $. Now in actual
applications, shear viscosity effects are often more important than either
bulk viscosity or heat conduction (cfr. ref. \cite{et}). We could therefore
retain only the third term, although that would mean our $Q$ would be merely
non negative, rather than positive definite.

We shall appeal to the ''$2\gg 1$ approximation'' (see Appendix) to retain
only the first of the two new terms in the thermodynamic potential, eq. (\ref
{tct2}), obtaining

\begin{equation}
\Phi ^a=\Phi _E^a+T^3u^a\left( 2Q+T\frac{\partial Q}{\partial T}\right)
\label{fs1}
\end{equation}

where the Eckart potential is still given by eq. (\ref{dt5b}). Assuming for
simplicity that the $\tau $'s are independent of $\alpha $ (we shall of
course assume that they are simply constant) the new addition to the
thermodynamic potential does not affect the expression for the particle
current in terms of field variables. The energy momentum tensor gets a new
term, quadratic in the $\zeta ^{ab}$'s, but we know that this term is small
and can be neglected, at least in a first run (see below). The real change
comes in the non equilibrium current $A^{abc}$, which now reads

\begin{equation}
A^{abc}=A_E^{abc}+T^3u^a\left[ \frac{4\tau _{BV}}{3\zeta _{BV}}\left(
P_{ST}\zeta \right) ^{bc}-\frac{\tau _{HC}}{2\kappa T}\left( P_V\zeta
\right) ^{bc}+\frac{\tau _{SV}}\eta \left( P_{TT}\zeta \right) ^{bc}\right]
\label{fs2}
\end{equation}

where $A_E^{abc}$ is the Eckart result eq. (\ref{nt1}). Its divergence
becomes

\begin{equation}
A_{;a}^{abc}=A_{E;a}^{abc}+T^3\left[ \frac{4\tau _{BV}}{3\zeta _{BV}}\left(
P_{ST}\dot \zeta \right) ^{bc}-\frac{\tau _{HC}}{2\kappa T}\left( P_V\dot 
\zeta \right) ^{bc}+\frac{\tau _{SV}}\eta \left( P_{TT}\dot \zeta \right)
^{bc}\right] +R^{bc}  \label{fs3}
\end{equation}

where $\dot \zeta _{de}=u^a\zeta _{de;a}$. The Eckart term is given in eq. (%
\ref{nt2}), and it is independent of the $\zeta $'s. The remainder $R^{bc}$
is a complicated expression involving products of $\zeta ^{ab}$'s and
derivatives of the temperature and four velocity; it is therefore of second
order in departure from equilibrium and can be neglected within present
accuracy.

As a consequence of Eq. (\ref{fs3}), the third set of equations of motion is
no longer merely algebraic; it now reads (cfr. eqs. (\ref{nt4} and \ref{st5}%
))

\begin{eqnarray}
&&A_{E;a}^{abc}+T^3\left[ \frac{4\tau _{BV}}{3\zeta _{BV}}\left( P_{ST}\dot 
\zeta \right) ^{bc}-\frac{\tau _{HC}}{2\kappa T}\left( P_V\dot \zeta \right)
^{bc}+\frac{\tau _{SV}}\eta \left( P_{TT}\dot \zeta \right) ^{bc}\right] 
\nonumber  \label{fs4} \\
&=&-T^3\left[ \frac 4{3\zeta _{BV}}\left( P_{ST}\zeta \right) ^{bc}-\frac 1{%
2\kappa T}\left( P_V\zeta \right) ^{bc}+\frac 1\eta \left( P_{TT}\zeta
\right) ^{bc}\right] +j^{bc}  \label{fs4}
\end{eqnarray}

Using the orthogonality properties of the $P$'s it is possible to decouple
these equations; the simplest case is when all $\tau $'s are equal, whereby
we simply get

\begin{equation}
\tau \dot \zeta ^{ab}+\zeta ^{ab}=\left( \frac{-1}{T^3}\right) \left[ \frac{%
3\zeta _{BV}}4P_{ST}^{abcd}-2\kappa TP_V^{abcd}+\eta P_{TT}^{abcd}\right]
\left\{ A_{Ecd;e}^e-j_{cd}\right\}  \label{fs5}
\end{equation}

We may also use eqs. (\ref{dt7}) and (\ref{nt2}) to transform eqs. (\ref{fs4}%
) into a set of Maxwell - Cattaneo equations for the viscous stresses\cite
{rmp}. In any case, the $\tau $'s represent the characteristic times in
which the $\zeta ^{ab}$'s (and therefore the viscous stresses as well) relax
to their Eckart values. The apparition of new coefficients with the physical
meaning of relaxation times is also characteristic of the ETT approach (cfr. 
\cite{jou,isrste80}) and to this extent we may conjecture that both
approaches are physically equivalent within this accuracy.

We can see now how the theory works. Because the $I^{ab}$ are the same, the
noise statistics has not changed, and in particular we still have Gaussian
white noise. The energy momentum tensor has not changed either, at least as
regards the expression of the viscous stresses in terms of the $\zeta _{ab}$%
's (there is a systematic contribution coming from the nonvanishing
expectation value of the new quadratic terms which ought to be included,
though). What has changed is that, if the noise - noise correlation is
ultralocal, then the stress - stress correlation cannot be, because the
solution to eq. (\ref{fs4}) will be necessarily non local in the noises. Of
course, we expect these correlations will decay exponentially on a time of
order $\tau ,$ and in the approximation in which these characteristic times
are neglected, we recover the Landau - Lifshitz results above.

\section{Acknowledgments}

This work is a spin-off from a larger project involving Oscar Reula and
Gabriel Nagy. It is a pleasure to acknowledge the comments and suggestions
received from both of them.

This work has been partially supported by the European project
CI1-CT94-0004, CONICET, UBA and Fundaci\'on Antorchas

\section{Appendix}

In this appendix. we shall take a closer view into the approximations
leading to Eq. (\ref{fs1}). The starting point is the generating functional
(cfr. Eq. (\ref{tct1}))

\begin{equation}
\chi =\chi _E+\gamma \left( T\right) Q\left( T,u^a,\zeta ^{ab}\right)
\label{tct1p}
\end{equation}

where $Q$ is a positive definite scalar quadratic form on the $\zeta ^{ab}$.
Then the thermodynamic potential (cfr. Eq. (\ref{tct2}))

\begin{equation}
\Phi ^a=\Phi _E^a+u^aT^2\frac d{dT}\left( \gamma Q\right) +\Delta
^{ab}T\gamma \frac{\partial Q}{\partial u^b}  \label{tct2p}
\end{equation}

If $\gamma =T^2$ and $Q$ is given by  Eq. (\ref{fs0}), we get

\begin{equation}
\Phi ^a=\Phi _E^a+\left( \frac{u^aT^3}2\right) \left[ \frac{4\tau _{BV}}{%
3\zeta _{BV}}\left( \zeta P_{ST}\zeta \right) -\frac{\tau _{HC}}{2\kappa T}%
\left( \zeta P_V\zeta \right) +\frac{\tau _{SV}}\eta \left( \zeta
P_{TT}\zeta \right) \right] +\Delta ^{ab}T\gamma \frac{\partial Q}{\partial
u^b}  \label{ctp1p}
\end{equation}

Recall also Eqs. (\ref{st6}, \ref{e11} and \ref{e12}).

The basic idea is that the timelike second term will dominate the spacelike
third term, thus ensuring negativity of $M_{BC}^aw_a$, where $w_a$ is an
arbitrary future oriented timelike vector, and $M_{BC}^a$ is the Hessian Eq.
(\ref{cdtt1}). If this is true, then the third term in Eq. (\ref{ctp1p}) may
be discarded. Since the second term is larger than the third roughly by a
factor $2$, we have called this the ''$2\gg 1$ approximation''.

(This argument does not hold for the heat conduction terms, which in fact
are linear in $T$; for them, we must rely directly on the argument below, or
else assume that $\tau _{HC}$ has a hidden temperature dependence.)

Of course, it is not hard to compute the third term explicitly, but is is
not very illuminating either. Our real interest lies in finding the tensor $%
A^{abc}$ and its divergence; when we do this, we find that the terms
proportional to $u^a$ contribute time derivatives of $\zeta ^{ab}$ as well
as several extra terms, called $R^{ab}$ in the text, Eq. (\ref{fs3}), while
the third term will only contribute space derivatives and cross terms
containing both the tensor $\zeta ^{ab}$ and derivatives of the temperature
and fluid velocity. Thus, in physical terms, the approximation involved is
that evolution in time is more important than inhomogeneity in space. In the
end we recover essentially the ETT equations of motion, which have their own
physical motivation, independent of the formal procedure above.

\end{document}